\documentclass[conference]{IEEEtran}
\IEEEoverridecommandlockouts

\usepackage{cite}
\usepackage{amsmath,amssymb,amsfonts}
\usepackage{algorithmic}
\usepackage{graphicx}
\usepackage{textcomp}
\usepackage{xcolor}
\usepackage{balance}
\usepackage{hyperref}
\usepackage{caption}
\usepackage{tabularx}
\usepackage{url}
\usepackage[most]{tcolorbox}
\tcbset{textmarker/.style={
        enhanced,
        parbox=false,boxrule=0mm,boxsep=0mm,arc=0mm,
        outer arc=0mm,left=5mm,right=3mm,top=7pt,bottom=7pt,
        toptitle=1mm,bottomtitle=1mm}}
\newtcolorbox{noteBox}{textmarker,
    borderline west={4pt}{0pt}{gray},
    colback=gray!10!white}
\newcommand{\note}[1]{\begin{noteBox} #1 \end{noteBox}}

\def\BibTeX{{\rm B\kern-.05em{\sc i\kern-.025em b}\kern-.08em
    T\kern-.1667em\lower.7ex\hbox{E}\kern-.125emX}}

\newcommand{\orcidlink}[1]{\textsuperscript{\href{https://orcid.org/#1}{\includegraphics[scale=0.2]{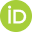}}}}
    
\begin{document}

\title{A Therapeutic Role-Playing VR Game for Children with Intellectual Disabilities}

\author{\IEEEauthorblockN{Santiago Berrezueta-Guzman\orcidlink{0000-0001-5559-2056}}
\IEEEauthorblockA{
\textit{Technical University of Munich}\\
Heilbronn, Germany \\
s.berrezueta@tum.de}
\and
\IEEEauthorblockN{WenChun Chen\orcidlink{0009-0007-9681-543X}}
\IEEEauthorblockA{
\textit{Technical University of Munich}\\
Heilbronn, Germany \\
wenchun.chen@tum.de}
\and
\IEEEauthorblockN{Stefan Wagner\orcidlink{0000-0002-5256-8429}}
\IEEEauthorblockA{
\textit{Technical University of Munich}\\
Heilbronn, Germany \\
stefan.wagner@tum.de}
}

\maketitle

\begin{abstract}

Virtual Reality (VR) offers promising avenues for innovative therapeutic interventions in populations with intellectual disabilities (ID). This paper presents the design, development, and evaluation of \textit{Space Exodus}, a novel VR-based role-playing game specifically tailored for children with ID. By integrating immersive gameplay with therapeutic task design, \textit{Space Exodus} aims to enhance concentration, cognitive processing, and fine motor skills through structured hand-eye coordination exercises. A six-week pre-test/post-test study was conducted with 16 children in Ecuador, using standardized assessments—the Toulouse-Pieron Cancellation Test and the Moss Attention Rating Scale—complemented by detailed observational metrics. Quantitative results indicate statistically significant improvements in concentration scores, with test scores increasing from 65.2 to 80.3 and 55.4 to 68.7, respectively (p $<$ 0.01). Qualitative observations revealed reduced task attempts, enhanced user confidence, and increased active participation. The inclusion of a VR assistant provided consistent guidance that further boosted engagement. These findings demonstrate the potential of immersive, game-based learning environments as practical therapeutic tools, laying a robust foundation for developing inclusive and adaptive rehabilitation strategies for children with ID.

\end{abstract}

\begin{IEEEkeywords}
Virtual Reality, Intellectual Disability, Therapeutic Gaming, Cognitive Enhancement, Hand-Eye Coordination, Human-Centered Design, Adaptive Learning Systems, Immersive Technology, Role-Playing Games, Rehabilitation Strategies, Therapeutic Games.
\end{IEEEkeywords}

\section{Introduction}\label{I}

Intellectual disability (ID) is a heterogeneous group of neurodevelopmental conditions characterized by significant limitations in cognitive functioning and adaptive behavior, such as daily living skills, social participation, and communication \cite{pratt2007intellectual}. Diagnosed during the developmental period (typically 0–18 years), ID is defined by an intelligence quotient (IQ) of less than 70 and impairments in adaptive skills, assessed through standardized measures like the Vineland Adaptive Behavior Scales \cite{estabillo2018adaptive}. ID affects approximately 2--3\% of children in high-income countries, with higher prevalence in low- and middle-income countries due to factors such as undernutrition, insufficient home stimulation, and infections. It can result from genetic causes, such as Down syndrome, or non-genetic factors, including environmental risks \cite{WHO_ICD11, parmenter2011intellectual}.

Children with ID are at an elevated risk for co-occurring mental health disorders, with prevalence rates that are at least double those of children without ID. Around 40\% of children with ID meet the criteria for a diagnosable mental disorder, highlighting the importance of targeted interventions and access to mental health services \cite{pratt2007intellectual}. Improved evidence-based approaches and equitable access to care are critical to addressing the unmet needs of children with ID \cite{totsika2022mental}. 

Virtual Reality (VR) is an immersive technology that creates simulated environments, enabling users to interact with and experience a digitally constructed world as if it were real. By employing head-mounted displays, motion controllers, and sensors, VR delivers a multisensory experience that combines visual, auditory, and tactile stimuli \cite{lavalle2023virtual}. This technology is used across various domains, from gaming and entertainment to education, healthcare, and training, allowing users to explore and engage with environments or scenarios that might be inaccessible or impractical in the physical world \cite{damianova2025serious}. VR relies on real-time rendering, user interaction, and spatial awareness to create a sense of presence. It is a transformative tool for experiential learning, therapeutic interventions, and creative expression \cite{Jerald2015, ShermanCraig2019}.

This paper presents the design, development, and evaluation of \textit{Space Exodus}\footnote{GitHub repository: https://github.com/TUM-HN/Space-Exodus}, a novel VR-based role-playing game specifically tailored for children with intellectual disabilities. Unlike prior work, our approach integrates immersive gameplay with therapeutic task design to enhance concentration and fine motor skills through structured hand-eye coordination exercises. To rigorously assess cognitive progress, we detail a pre-/post-test study conducted with 16 children in Ecuador, incorporating both behavioral and observational metrics, including the Toulouse and Pieron Cancellation Test and the Moss Attention Rating Scale. 

This paper is organized as follows: Section~\ref{RW} reviews existing applications of Virtual Reality for individuals with intellectual disabilities. Section~\ref{M} presents the methodology, including participant demographics, task design, and experimental protocol. Section~\ref{R} describes the results of the intervention, supported by both quantitative data and qualitative observations. Section~\ref{Ch} outlines the main challenges encountered during the study. Section~\ref{D} discusses behavioral insights and game design implications. Finally, Section~\ref{C} concludes the paper.

\section{Related Work}\label{RW}

The application of Virtual Reality (VR) in supporting individuals with Intellectual Disabilities (ID) has gained increasing attention in recent years. Researchers have explored VR’s potential to enhance physical activity, develop life skills, and provide therapeutic interventions tailored to the needs of individuals with ID. 

McMahon et al. \cite{mcmahon2020virtual} investigated using virtual reality (VR) exergaming to address physical activity challenges among high school students with intellectual and developmental disabilities (IDD). The study aimed to increase exercise duration and intensity using immersive VR games, involving four participants in a single-subject multiple probe design. Results demonstrated that all participants significantly improved their exercise time and calorie expenditure during VR exergaming sessions compared to baseline conditions. The intervention also had high social validity, with students and teachers reporting positive experiences. Despite its promising outcomes, the study highlighted limitations, such as the high initial equipment cost and the need for setup assistance. Future research directions include exploring long-term effects, broader participant samples, and alternative immersive exercise technologies.

Cheung et al., \cite{cheung2022virtual} conducted a multicenter randomized controlled trial to evaluate the effectiveness of a Virtual Reality (VR)-based life skills training program for individuals with ID. The study compared VR training to traditional and control groups, focusing on grocery shopping, cooking, and kitchen cleaning tasks. Results demonstrated significant improvements in the VR group's cooking and cleaning performance and memory span compared to controls. However, traditional training showed greater effectiveness for shopping tasks. The study highlights VR’s potential to create engaging and secure training environments. However, it faced limitations such as sample size constraints, a lack of standardized assessment tools, and the challenge of customizing training for varying IQ levels. Future directions emphasized integrating artificial intelligence for personalized training and addressing the long-term efficacy of VR-based interventions.

Franze et al. (2024) conducted a study comparing the effectiveness of immersive virtual reality (VR) headsets to non-immersive devices in teaching practical life skills to adults with intellectual disabilities. The study involved 36 participants who underwent 12 virtual training sessions focused on waste management tasks, such as sorting general waste from recycling and organics. Findings indicated that the group using immersive VR headsets demonstrated significantly better real-world performance in correctly sorting waste immediately after training and maintained this improvement up to a week later, compared to the non-immersive group. The study suggests that immersive VR can provide realistic, hands-on learning experiences in a safe environment, potentially enhancing the acquisition of life skills in this population. However, considerations such as the initial cost of VR equipment and the need for setup assistance were noted as potential limitations \cite{franze2024immersive}.

Cortés et al. (2024) developed an immersive behavioral therapy tool tailored for individuals with ID to address phobias, focusing on stair-related anxiety. Integrating systematic desensitization techniques with eXtended Reality (XR), the tool combines 360-degree videos and 3D virtual environments to simulate anxiety-provoking situations in a controlled setting. Therapists monitor real-time biomarker data, such as heart rate, through wearable devices, enabling dynamic adjustments to the therapy session. This system incorporates advanced body visualization using neural networks, offering a natural and engaging user experience. While the study demonstrated the potential of XR in enhancing behavioral therapy for individuals with ID, limitations included the reliance on external hardware and the need for therapist guidance. Future work aims to streamline hardware integration and expand its applicability to other phobias \cite{cortes2024immersive}.

Given the routine and monotonous nature of daily rehabilitation for ID children, the limitations of outdoor physical activities due to safety concerns, and the need for high caregiver-to-child ratios, VR technology offers a promising solution. By simulating various environments, VR can enhance external stimuli, provide desensitization therapy, and alleviate anxiety associated with unfamiliar external environments (i.e., environments beyond their daily living spaces, such as care facilities, homes, and medical institutions). Furthermore, VR can facilitate extensive practice for specific skills. As demonstrated by Cherix et al., VR-simulated road crossing training can address the safety concerns of real-world scenarios while simultaneously developing essential self-care skills \cite{cherix2020training}. Their study utilized VR to simulate diverse road crossing scenarios, enabling ID children to practice and improve their judgment in a safe environment. In addition, physical activity has been shown to improve children's concentration abilities \cite{korkusuz2023does}. 

A previous study introduced \textit{Space Exodus}, a task-based role-playing VR game to enhance therapeutic interventions for children with ID. The game immerses users in everyday scenarios to facilitate skill acquisition and transfer, with preliminary experiments indicating that 70--80\% of participants successfully applied learned skills to new challenges. However, the study identified limitations, including VR-induced discomfort, controller misoperation, and task complexity. This underscores the need for ergonomic improvements and adaptive guidance to optimize the user experience and therapeutic outcomes \cite{chen2024task}.

\section{Methodology}\label{M}

This paper aims to develop a Role-Playing Game (RPG) set in a space environment, incorporating a series of tasks that simulate daily life situations. By integrating numerous hand-eye coordination exercises into the game, we provide children with engaging and enjoyable training that can enhance their concentration skills through fine motor skill practice. This methodology section discusses the selection of participants, the hardware and tasks involved in the study, and the experiment design and data analysis methods.

\subsection{Participants}

Our current study involves 16 children aged 8 to 12 diagnosed with ID. Based on the Diagnostic and Statistical Manual of Mental Disorders, Fifth Edition (DSM-5) criteria \cite{APA2013}, the participants are classified according to the severity of their condition as we illustrate it in Table \ref{tab:classification}, and according to their co-occurring conditions as we illustrate it in Table \ref{tab:cooccurring}.
Participants were recruited through local educational institutions and therapy centers that specialize in intellectual and developmental disabilities, in coordination with professionals and caregivers who provided informed consent.

\begin{table}[ht]
\caption{Classification of the participants according to DSM-5}
\centering
\begin{tabularx}{\columnwidth}{|X|c|}
\hline
\textbf{Classification} & \textbf{Participants} \\
\hline
Mild Intellectual Disability         & 5 \\
Moderate Intellectual Disability     & 4 \\
Severe Intellectual Disability       & 4 \\
Profound Intellectual Disability     & 3 \\
\hline
\end{tabularx}
\label{tab:classification}
\end{table}

Inclusion criteria required a formal diagnosis of intellectual disability based on DSM-5 standards. Exclusion criteria included severe visual or motor impairments that would prevent using VR equipment.

\begin{table}[ht]
\caption{Classification of the participant according to co-occurring conditions}
\centering
\begin{tabularx}{\columnwidth}{|X|c|}
\hline
\textbf{Condition} & \textbf{Participants} \\
\hline
ADHD (Attention-Deficit/Hyperactivity Disorder) & 4 \\
Chromosomal Disorder                            & 1 \\
Depression                                       & 1 \\
Anxiety                                          & 3 \\
ASD (Autism Spectrum Disorder)                  & 1 \\
Hearing Loss                                     & 1 \\
Cerebral Palsy (CP)                              & 2 \\
Seizures                                         & 1 \\
Muscular Dystrophy                               & 1 \\
Down Syndrome (DS)                               & 1 \\
\hline
\end{tabularx}
\label{tab:cooccurring}
\end{table}

The study protocol was approved by the specific center in the country of study (Anonymous during the review), and all procedures complied with international ethical standards for research involving human subjects.

\subsection{Hardware and Software}
In this paper, the game \textit{Space Exodus} is developed for the Meta Quest 2 and Meta Quest 3. These headsets were chosen due to their portability, wireless design, and affordability, making them suitable for educational or therapeutic settings with minimal infrastructure requirements. The headsets offer 6 degrees of freedom (6DoF) tracking and hand controller input, enabling precise hand-eye coordination tasks critical to the therapeutic goals of the game. Meta Quest 3 also provides improved passthrough resolution and processing power compared to its predecessor \cite{metaquest3docs}.

The engine used in this development is \textit{Unity}. This cross-platform game engine, developed by \textit{Unity Technologies}, is widely used for creating 2D, 3D, VR, and AR experiences. It provides an integrated development environment with real-time rendering capabilities, physics simulation, animation tools, and scripting via \textit{C\#}. Unity supports deployment to a broad range of platforms, including desktop, mobile, web, and XR devices, making it a popular choice for both game development and serious applications in education, simulation, and healthcare \cite{unitydocs}.

\subsection{Tasks Description}

The game consists of five tasks designed to progressively introduce players to the game mechanics and environment. 

\textit{Task 1 -- Throw the meteor into the trashcan} is an introductory practice, familiarizing players with character movement and object interaction using the hand controllers. 

\textit{Task 2 -- Clean up the mess in the bricks room} builds upon the skills learned in Task 1, requiring players to grab and place objects into corresponding color-coded boxes. This task simulates a familiar daily scenario, such as cleaning and recycling. 

\textit{Tasks 3 and 4 -- Disintegrate the larger meteor and put energy source on an analyzer} challenge players to use a gun to disintegrate meteors and then place the energy source onto an analyzer. These tasks further develop players' object manipulation skills and introduce more interactive elements, preparing them for later stages of the game. 

\textit{Task 5 -- Pilot spaceship to the wormhole}, players must navigate the game map, including climbing ladders, to reach the mission location. This segment tests players' hand-eye coordination and concentration. Additionally, the players pilot a spaceship through an asteroid field to get a wormhole, utilizing the skills and knowledge acquired throughout the previous tasks.

\subsection{Experiment Design}
This study did not utilize distinct experimental and control groups. Instead, a pre-test/post-test design was employed, supplemented by weekly assessments, to measure changes in concentration among the ID children. The total duration of the study is six weeks.

The Toulouse and Pieron Cancellation Test and the Moss Attention Rating Scale are our primary metrics for quantifying concentration levels. The Toulouse and Pieron Cancellation Test is a focused attention assessment suitable for individuals aged six through adulthood. It is widely used to evaluate attention span, visual search skills, and mental processing speed, particularly in tasks requiring sustained concentration over extended periods. This test is frequently employed in cognitive evaluations for both children and adults, especially in cases involving ADHD, brain injury, or neuropsychological research. The test takes approximately ten minutes, and results yield metrics for both speed and processing precision \cite{alberto2003atenccao}.

Recognizing that not all participants could complete the Toulouse and Pieron Cancellation Test, we incorporated the Moss Attention Rating Scale. This scale is a behavioral assessment typically completed by teachers, parents, or clinicians based on observations of an individual's behavior in everyday settings. While more commonly used in assessing patients with brain injuries, the Moss Attention Rating Scale is also employed to evaluate attention problems in individuals with ADHD \cite{whyte2003moss}, \cite{hart2006dimensions}. Given the specific causes of attention difficulties in our participants, this metric focuses on the change in scores throughout the study rather than the absolute score itself. The same version of the scale was used for each participant to ensure consistency.

In addition to measuring concentration ability, one month after the experiment, we compiled a questionnaire for observers to record the following: Daily and weekly game usage time.
Average time required for each task and any changes in time spent (if tasks were repeated). Obstacles encountered by each participant in the game
Occurrence of VR sickness. Any areas where the use of VR equipment was limited due to discomfort, such as dizziness or claustrophobia. Finally, some game design details will be provided as suggestions in the results section to improve the game design.

The experimental procedure is designed to observe the improvement in concentration ability through a VR game. For each participant, the process will involve a pre-test, gameplay sessions, mid-term (week 3) follow-up tests, and a post-test at the end of the experiment.

\begin{figure}[htbp]
\centerline{\includegraphics[width=9cm]{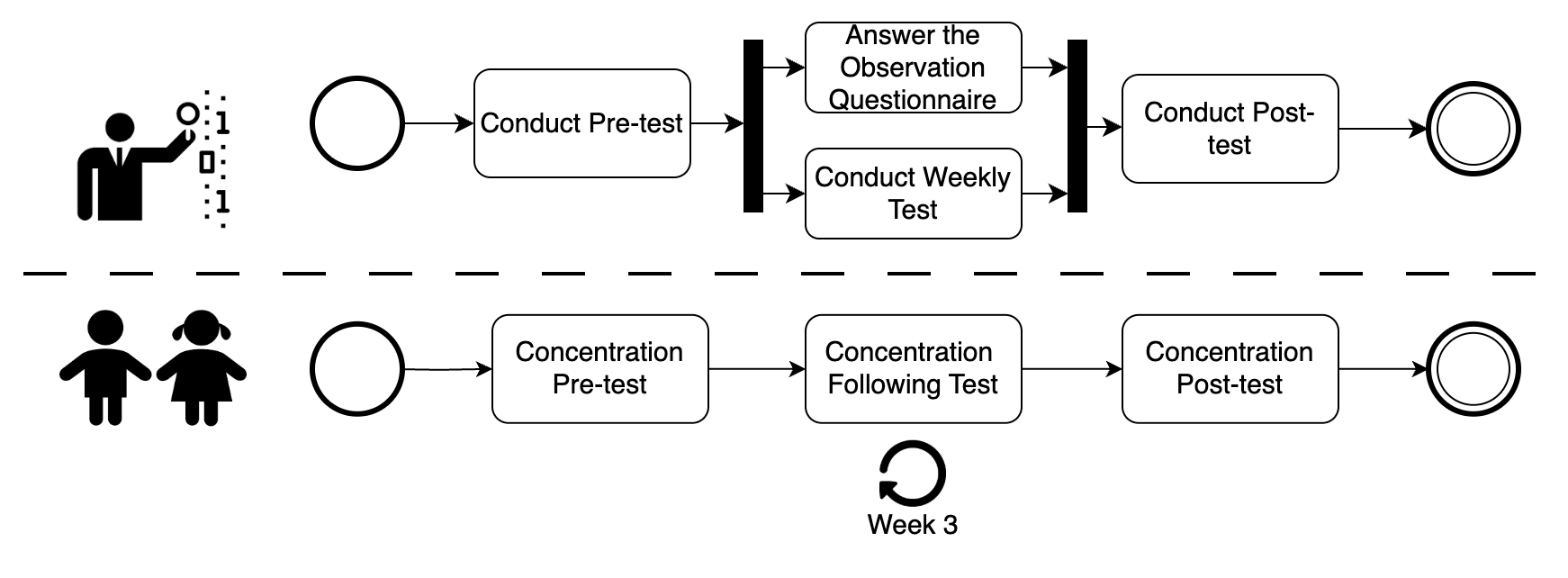}}
\captionsetup{font=normalsize}
\caption{Experimental Activity Flowchart. The experiment begins with a pre-test, followed by a mid-term assessment in Week 3, and concludes with a post-test in Week 6.}
\label{fig:expflowchart}
\end{figure}

Figure \ref{fig:expflowchart} shows the experiment steps. For the observers, the process will include conducting the pre-test, accompanying the participants during the initial gameplay sessions to observe any VR-related discomfort and provide immediate assistance, tracking gameplay time, and implementing recommended rest periods to prevent visual impairment caused by prolonged VR use (with a suggested break of 20-30 minutes each time, and maximum continuous use of two hours for VR equipment \cite{yoon2021effects}). One month after the start of the experiment, the observer will complete a game observation scale, provide statistical data on game duration, conduct a mid-term (week 3) follow-up test, and finally, assist in implementing the post-test at the end of the experiment.

As a side note, we specify that all headsets and controllers were sanitized between sessions, and disposable face covers were used to maintain hygiene and reduce the risk of infection transmission. Gameplay was conducted in quiet, well-lit rooms with soft flooring and ample space for movement. The area was cleared of obstacles to minimize the risk of falls or disorientation during VR use.

\subsection{Data Analysis}
In this experimental study, we employed five observational measures to investigate various aspects of user experience with the VR game. These measures include the initial play and rest time distribution, sustained interest in the game, route planning time within the game environment, the correlation between specific adaptive deficiency domains and encountered challenges in different game levels, and the potential impact of the game design on ID concentration ability.

To understand the initial user experience and adaptation to VR technology, we analyzed the duration of play sessions and the intervals of rest time taken between sessions during the first week. This analysis aimed to determine when users need to overcome any initial discomfort or dizziness associated with the VR headset.

We tracked the variation in play session duration over several weeks to assess the long-term engagement with the RPG game format. In conjunction with a monthly questionnaire, this longitudinal data provided insights into the most appealing tasks, tasks that users were willing to repeat, and tasks that presented significant challenges requiring additional guidance.

By measuring the route planning time skill, we can evaluate the impact of VR environment exploration on spatial awareness. Based on existing literature, we hypothesized that navigating within the VR environment would aid in map construction and enhance users' sense of direction, potentially benefiting their real-world orientation \cite{kober2013virtual}. Therefore, the game was intentionally designed without an in-game map. By recording the time taken to locate mission objectives, we aimed to assess the efficiency of route planning and whether it improved with task progression.

Finally, we analyzed the relationship between adaptive deficiency domains and the specific challenges users encounter at different game levels. Additionally, we conducted a t-test to examine the improvement in concentration ability over time.

This comprehensive approach, utilizing both quantitative and qualitative data, allowed us to better understand the user experience and the potential benefits of VR intervention in rehabilitation.

\section{Results}\label{R}

This section presents the outcomes of our six-week study on 16 children with intellectual disabilities. We assessed concentration improvements using the Toulouse and Pieron Cancellation Test and the Moss Attention Rating Scale and monitored task performance within the \textit{Space Exodus} game.

\subsection{Progression of Task Completion}

\begin{figure*}
    \centering
    \includegraphics[width=\linewidth]{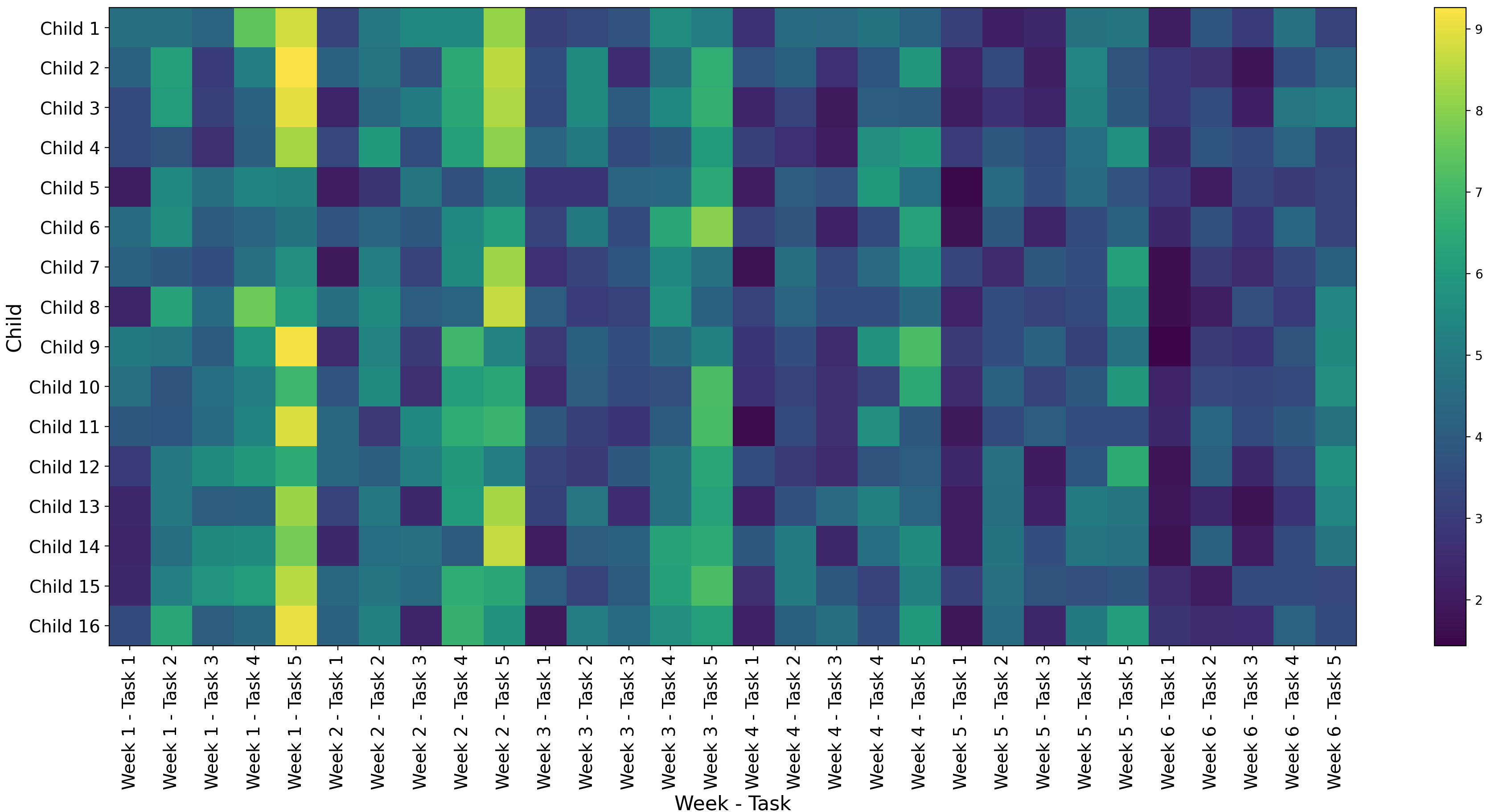}
    \caption{Progression of task completion times for 16 children over six weeks. Each column represents a specific task (Week–Task), and each row corresponds to a child. Color gradient indicates completion time, with lighter shades representing longer durations and darker shades representing shorter durations.}
    \label{fig:heatmap}
\end{figure*}

Figure~\ref{fig:heatmap} illustrates the progression of task completion times over the study period. On average, the total time to complete the five tasks decreased from 35 minutes in Week~1 to 22 minutes in Week~6, reflecting an apparent learning effect. This is evident because the color gradient shifts from lighter (indicating longer completion times) to darker shades (indicating shorter completion times) over the six weeks. 

However, specific tasks, particularly \textit{Task 5} in the early weeks, show lighter shades (yellowish), meaning they took longer to complete initially but improved notably in later weeks.
Additionally, some children (e.g., Child 1, Child 3, Child 9, and Child 15) initially required more time than their peers, especially in complex tasks (yellow regions), but also demonstrated considerable improvement.
By week 6, most children consistently improved (darker-colored) performances, indicating that they learned and retained skills effectively through repeated exposure and practice.

\subsection{Concentration Assessments}

\begin{figure}[htbp]
\centerline{\includegraphics[width=9cm]{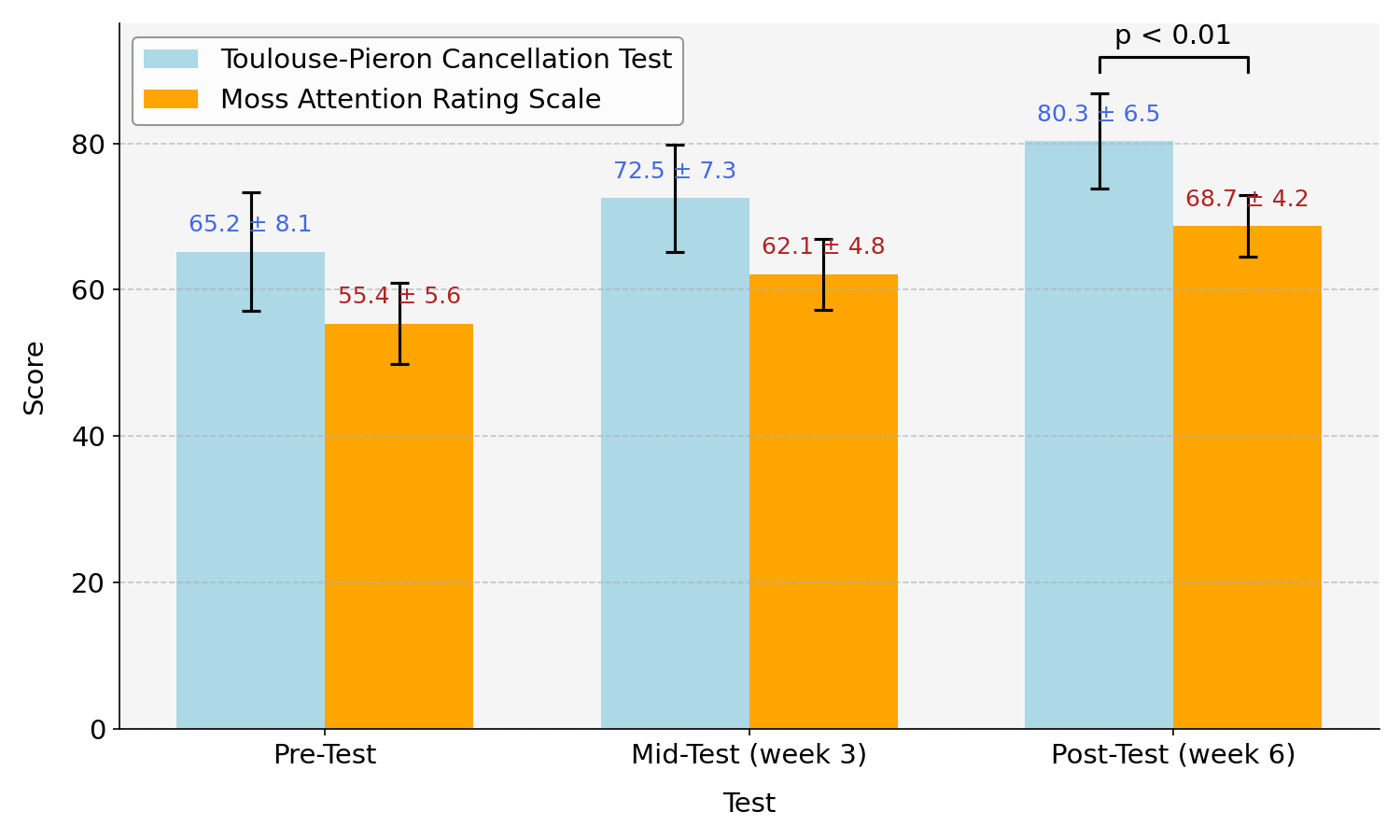}}
\captionsetup{font=normalsize}
\caption{Mean concentration assessment scores across Pre-Test, Mid-Test, and Post-Test intervals.}
\label{fig:Bars}
\end{figure}

Table~\ref{tab:concentration} summarizes the mean scores (with standard deviations) for the concentration assessments administered at three time points: pre-test, mid-test (Week~3), and post-test (Week~6). The results indicate steady improvements in both attention and processing speed, as we illustrate in Figure \ref{fig:Bars}.

\begin{table*}[t]
\caption{Mean Concentration Assessment Scores}
\centering
\begin{tabular}{|l|c|c|c|}
\hline
\textbf{Assessment} & \textbf{Pre-Test (Mean ± SD)} & \textbf{Mid-Test (Mean ± SD)} & \textbf{Post-Test (Mean ± SD)} \\
\hline
Toulouse-Pieron Cancellation Test&65.2 $\pm$ 8.1&72.5 $\pm$ 7.3&80.3 $\pm$ 6.5 \\
Moss Attention Rating Scale          & 55.4 $\pm$ 5.6 & 62.1 $\pm$ 4.8 & 68.7 $\pm$ 4.2 \\
\hline
\end{tabular}
\label{tab:concentration}
\end{table*}

A paired t-test confirmed that the improvements from the pre-test to the post-test were statistically significant with a \(p < 0.01\) for both assessments, supporting the efficacy of the VR intervention.

\subsection{Qualitative Observations}

Observational data provided further insights into the user experience. In the initial sessions, children required an average of 3.5 attempts per task; by Week~6, this number decreased to 1.8, indicating improved task comprehension and motor coordination. 

One caregiver commented, “She no longer waits for me to tell her what to do—she just picks up the controller and starts the task on her own.”
Similarly, a teacher noted improved task understanding: “By the third week, most of the children were completing the missions without needing me to repeat the instructions.”

Group participation also increased significantly, rising from 40~\% at the beginning to 75~\% by the end of the study. One facilitator said, “For the first time, I saw him raising his hand to try the next level. That has never happened before.”

\subsection{Severity and Co-occurring Conditions}

To further analyze the efficacy and applicability of the \textit{Space Exodus} intervention, we illustrated in Table \ref{tab:additional_analysis} differences in outcomes based on the classification of intellectual disability and co-occurring conditions.

\begin{table}[htbp]
\caption{Relationship between ID severity, co-occurring conditions, and concentration improvement}
\centering
\resizebox{\columnwidth}{!}{%
\begin{tabular}{|l|c|c|c|}
\hline
\textbf{Participant Group} & \textbf{Avg. Pre-test Score} & \textbf{Avg. Post-test Score} & \textbf{Improvement (\%)} \\
\hline
\multicolumn{4}{|c|}{\textbf{Intellectual Disability Severity}} \\
\hline
Mild (n=5) & 68.1 & 83.5 & 22.6 \% \\
Moderate (n=4) & 65.5 & 80.7 & 23.2 \% \\
Severe (n=4) & 62.9 & 78.0 & 24.0 \% \\
Profound (n=3) & 61.8 & 75.9 & 22.8 \% \\
\hline
\multicolumn{4}{|c|}{\textbf{Co-occurring Conditions}} \\
\hline
ADHD (n=4) & 60.7 & 77.4 & 27.5 \% \\
Anxiety (n=3) & 63.5 & 78.9 & 24.3 \% \\
Cerebral Palsy (n=2) & 58.4 & 72.3 & 23.8 \% \\
Other conditions (n=7) & 66.2 & 81.1 & 22.5 \% \\
\hline
\end{tabular}%
}
\label{tab:additional_analysis}
\end{table}

These differentiated findings underscore the importance of tailoring VR therapeutic interventions based on individual cognitive and co-occurring conditions, suggesting potential benefits from adaptive game mechanics and personalized engagement strategies in future iterations of the \textit{Space Exodus} platform.

\subsection{User Feedback}

Feedback collected via observer questionnaires highlighted the positive reception of the VR experience. Approximately 85\% of participants expressed enjoyment, with many citing the immersive and interactive nature of the game as key motivators. In addition, 90\% of caregivers observed enhanced concentration and a more remarkable ability to follow multi-step instructions during daily activities.

Overall, the quantitative improvements and qualitative observations demonstrate that \textit{Space Exodus} significantly enhances concentration and cognitive engagement in children with intellectual disabilities, affirming its potential as an effective therapeutic tool.

\subsection{Observations}

During this study, we observed that...

\begin{itemize}
    \item Tasks that involve sequencing, spatial reasoning, or coordination (e.g., aligning objects, navigating) are complicated for those with conceptual deficits.
    \item Tasks requiring physical action or movement (e.g., climbing ladders, spaceship navigation) are significant barriers for those with practical domain deficiencies.
    \item Tasks dependent on following social rules or understanding instructions pose challenges for those with social domain deficiencies exacerbated by anxiety or ADHD.
\end{itemize}

\section{Challenges}\label{Ch}

\subsection{VR Usage Challenges}

A significant challenge encountered during the study was VR sickness, affecting approximately 25\% of participants. Symptoms such as dizziness and slight nausea were common in the initial sessions but gradually diminished after one or two sessions at 20-minute intervals. Additionally, the weight and fit of the VR headset caused discomfort for some children, leading them to frequently adjust the device. Another common issue was the children's limited understanding of VR controls. Many initially attempted to turn their bodies instead of using the controllers and struggled with grasping mechanics, which led to frustration when interacting with objects. Furthermore, fatigue also played a role in engagement levels. While the average session lasted between 20 and 30 minutes, aligning with recommended breaks, some children exhibited exhaustion due to prolonged VR use's cognitive and physical demands.

\subsection{Game Interaction Challenges}

Children's ability to complete the tasks within the game varied significantly. Task 1, which involved throwing a meteor, was the easiest and quickest to complete, typically requiring around two to four minutes. On the other hand, Task 5, where children piloted a spaceship, took the longest to finish, averaging six to ten minutes due to its dynamic and interactive nature. 

The total game completion time ranged from 20 to 35 minutes, with a variance of approximately five minutes depending on each child's level of engagement and comfort. Among all the tasks, piloting the spaceship was the most engaging, while cleaning up the mess in Task 2 was perceived as repetitive and less stimulating. Tasks that require complex mechanics, such as disintegrating the meteor or placing the energy source, had the highest failure rates. Children typically make three to four attempts before becoming visibly frustrated or disengaged when encountering difficulties.

Navigation within the game environment also presented challenges. While 70\% of participants focused on solving the tasks, about 30\% spent time exploring the spaceship environment. Locating mission areas was particularly difficult for some children, with the brick room taking around three to five minutes to find and the control room around two to three minutes. The absence of an in-game map prolonged exploration times for some players. Additionally, voice instructions were not fully adequate, as all participants required additional explanations from teachers in Spanish to understand the game objectives. The story component of the game also did not capture the children's attention, with educators summarizing key points for them in simple phrases.

\section{Discussion}\label{D}

The results of our study demonstrate that the \textit{Space Exodus} VR intervention yields substantial improvements in both behavioral and cognitive performance among children with intellectual disabilities. Among the findings we have: 

\note{\textbf{Finding 1: Measurable gains in concentration and cognitive focus.} The study demonstrated that children with intellectual disabilities experienced significant improvements in concentration after engaging with the Space Exodus VR intervention over six weeks.} The Toulouse-Pieron Cancellation Test scores rose from 65.2 to 80.3, while the Moss Attention Rating Scale increased from 55.4 to 68.7. These improvements were statistically significant \((p < 0.01)\), underscoring the intervention’s effectiveness in enhancing sustained attention and cognitive processing through immersive gameplay. 

While Cheung et al. \cite{cheung2022virtual} reported VR-based life skills training improved performance in controlled tasks, they noted traditional methods were more effective in some areas. Our findings suggest that, when well-designed and customized, VR can outperform conventional methods in enhancing attention, especially with children showing co-occurring conditions such as ADHD, where we observed the highest improvement (27.5\%).

Franze et al. \cite{franze2024immersive} demonstrated that immersive VR led to better task retention than non-immersive systems. Similarly, we found that children in our study improved task completion speed and required fewer attempts, indicating increased comprehension and memory consolidation.

\note{\textbf{Finding 2: Progressive mastery of tasks through repetition and engagement.}
Over the six-week program, participants showed evident progress in completing in-game tasks.} The average total time to finish all five tasks decreased from 35 minutes in Week 1 to 22 minutes in Week 6, and the average number of attempts per task fell from 3.5 to 1.8. This progression reflects growing familiarity with the VR environment, improved motor coordination, and better task comprehension. More challenging tasks, particularly those involving spatial navigation or complex interactions, showed the most notable gains over time.

\note{\textbf{Finding 3: Growth in confidence, participation, and social engagement}
The intervention also positively influenced participants’ behavior and self-confidence.} Active group participation rose from 40\% at the beginning to 75\% by the end of the study. Teachers and caregivers observed that the children became more self-directed, showed increased motivation to complete tasks, and relied less on adult guidance. This behavioral shift indicates immersive, task-oriented VR can foster cognitive, emotional, and social growth.

Compared to McMahon et al. \cite{mcmahon2020virtual}, who found that VR exergaming improved physical engagement, our work shows that immersive role-playing tasks can go beyond physical activity and substantially enhance cognitive performance, particularly sustained attention.

\note{\textbf{Finding 4: Consistent benefits across different disability severity levels and profiles.}
A breakdown of outcomes based on the severity of intellectual disability revealed consistent improvement across all groups, from mild to profound.} Children with co-occurring conditions, such as ADHD, anxiety, or cerebral palsy, also showed substantial gains, with those in the ADHD group achieving the highest improvement (27.5\%). These results highlight the adaptability of the game design to various cognitive and behavioral needs, reinforcing the potential of personalized, inclusive therapeutic tools.

\note{\textbf{Finding 5: Critical support role of the integrated VR assistant.}
The embedded VR assistant was a key feature in facilitating the children's interaction with the game.} By offering consistent, patient, and non-intrusive guidance, the assistant helped children understand complex instructions, reduce frustration, and feel more supported throughout the experience. This interactive component benefited participants who were less comfortable seeking help from human facilitators, enabling more autonomous exploration and learning.

\note{\textbf{Finding 6: Therapeutic potential of game-based, immersive interventions.}
The findings affirm that Space Exodus is an effective and engaging therapeutic medium for children with intellectual disabilities.} Its structured role-playing format, adaptive challenges, and engaging design supported cognitive improvements and positive behavioral shifts. The success of this intervention paves the way for further research into scalable, VR-based rehabilitation tools that are inclusive, motivating, and tailored to diverse learner profiles.
These comparisons underline the value of immersive VR interventions that are contextually meaningful, goal-oriented, and supportive through features like the virtual assistant used in Space Exodus.

\subsection{Limitations and Future Work}

The relatively small sample size of 16 participants limits the generalizability of our results. Furthermore, the study's six weeks may not fully capture the long-term retention of the observed improvements. Future work should include more diverse populations and extend the observation period to assess the persistence of these gains. Additionally, incorporating adaptive algorithms to personalize task difficulty in real-time could optimize the intervention for individual needs.

This study's improved cognitive and behavioral outcomes suggest that \textit{Space Exodus} is a viable therapeutic tool. Integrating immersive VR technology with structured, task-based learning can significantly enhance concentration and motor coordination in children with intellectual disabilities, paving the way for more innovative and inclusive rehabilitation strategies. Furthermore, the platform can be extended by incorporating real-life objects into the virtual environment using photogrammetry, creating more realistic and relatable scenarios. Such enhancements can increase user engagement by simulating familiar environments, as demonstrated in our prior work on virtual world construction through high-fidelity 3D modeling techniques based on photogrammetric data \cite{berrezueta2025reality}. This opens up opportunities for personalized therapeutic experiences that better reflect users’ daily needs.

\subsection{Implications for Design and Practice}

The structured progression of tasks, visual clarity, and consistency of feedback were instrumental in sustaining engagement. Our results reinforce the importance of user-centered design in therapeutic VR, echoing Cherix et al. \cite{cherix2020training}, who emphasized safety and task realism in virtual environments. 

The success of our VR assistant highlights the need for embedded guidance tools that reduce frustration and foster autonomy, particularly in children with social or communication challenges. This aligns with findings from Cortés et al., \cite{cortes2024immersive}, who also reported benefits from guided support in XR environments.
Finally, the variation in performance by severity and co-occurring condition supports calls from previous studies for adaptive systems that personalize difficulty and real-time feedback.

\section{Conclusion}\label{C}

This paper presented Space Exodus, a role-playing VR game designed to enhance attention and fine motor skills in children with intellectual disabilities. Through a six-week intervention with 16 participants, we observed statistically significant improvements in concentration (Toulouse-Pieron: 65.2 to 80.3; Moss Scale: 55.4 to 68.7) and reductions in task completion time and failure rates.

Unlike prior work focused primarily on physical activity or life skills, Space Exodus integrates therapeutic design into an engaging narrative-driven experience supported by a virtual assistant. The positive feedback from caregivers and observable behavioral gains confirm the potential of immersive VR in inclusive rehabilitation.
Future efforts should focus on personalization, scalability, and longer-term evaluation. Our findings lay the groundwork for broader adoption of VR interventions in cognitive therapy and inclusive education.

\balance
\bibliographystyle{ieeetr}
\bibliography{RolePlayGame}

\end{document}